\pdfoutput=1

\documentclass[sts,preprint]{imsart}

\usepackage{amsthm,amsmath,natbib}
\RequirePackage[colorlinks,citecolor=blue,urlcolor=blue]{hyperref}

\startlocaldefs
\let\code=\texttt
\let\pkg=\texttt
\endlocaldefs

\begin{document}

\begin{frontmatter}

\title{Self-exciting Point Processes: Infections and Implementations}
\runtitle{Comment}

\author{\fnms{Sebastian} \snm{Meyer}\ead[label=e1]{seb.meyer@fau.de}}
\address{Sebastian Meyer is a Research Fellow at the Institute of Medical
  Informatics, Biometry, and Epidemiology, Friedrich-Alexander-Universit\"at
  Erlangen-N\"urnberg, 91054 Erlangen, Germany \printead{e1}.}
\runauthor{S. Meyer}

\end{frontmatter}

Thanks for this overdue account of
\emph{self-exciting} spatio-temporal point process models,
synthesizing developments from various research fields.
In what follows, I will contribute some experiences from modelling the
spread of infectious diseases (relating to Section~4.3 of the review).
Furthermore, I will try to complement the review with regard to
the availability of software for the described models,
which I think is essential in ``paving the way for new uses''.

\section*{Point process models for infectious disease spread}

For notifiable diseases, public health surveillance data is routinely available
in aggregated form as time series of infection counts. Such
data are typically approached with autoregressive models using a negative
binomial distribution, or assuming the counts as approximately Gaussian after a
suitable transformation 
to adopt classical ARIMA models or even Facebook's Prophet
procedure \citep[see][for an assessment]{held.meyer2018}.
For \emph{multivariate} time series stratified by region, spatial
epidemic models can account for varying demographic and
environmental factors,
and enable spatially explicit predictions. \citet{hoehle2016} provides a
recent overview of spatio-temporal infectious disease models.

\citet{taylor.etal2015} propose to tackle even such aggregate-level
surveillance data with point process methods (specifically,
a log-Gaussian Cox process model with Bayesian data augmentation).
However, for ``mechanistic'', 
self-exciting point process models to
unfold in infectious disease epidemiology, individual-level data are
indispensable.
A distinction is between a point process indexed in a continuous
spatial domain, such as in the ETAS model, versus a multivariate
temporal point process operating on a discrete set of
interacting locations/individuals, i.e., on a network. Reinhart mentions
recent applications of such multivariate processes in social networks.
It is important to note that similar models have also
been developed in the infectious disease context, where they are not that
much ``in its infancy''. For instance, the models described in
\citet{diggle2006}, \citet{scheel.etal2007}, and \citet{hoehle2009}
all describe the spread of livestock diseases among
farms using distance-based transmission kernels. Such spatial distances
could just as well be replaced by geodesic distances to quantify the
coupling between the individual infection processes, for example using movement
networks as in \citet{schroedle.etal2012} or contact networks as mentioned
by Reinhart. \citet{aldrin.etal2015} use a combination of spatial
distances and local contact networks.

In what follows, I focus on spatially \emph{continuous} self-exciting
point process models for the spread of infectious diseases in human
populations. Such models come with several caveats, on three of which I
would like to elaborate.

\subsection*{Limited spatio-temporal data resolution}

The available spatial resolution of case reports is often
limited by data protection. This constrains the detail with which
spatial interaction can be estimated. ``Areal censoring'' (e.g., to the
postcode level) may yield events that apparently occurred at the same
location, which is impossible in simple point processes. Equivalently, interval
censoring of the infection times results in concurrently observed events,
making it impossible to ascertain which
infection predates the other. Furthermore, the
situation is complicated by the fact that event times only correspond to
the date of specimen sampling or notification to public health
authorities. As latent periods 
and reporting delays differ between cases,
the observed ordering of the events may not always properly reflect the
infection chain.

One way of dealing with tied event times and locations is to
add random jitter with an amount corresponding to the level of censoring in the
data, and ideally conduct a sensitivity analysis or use model averaging over
several random seeds. Breaking ties will affect estimates of the triggering
function as well as it will remove
spikes in the distribution of rescaled temporal residuals
\citep[see][Figure~4]{meyer.etal2012}. These are described in
\citet[Section~3.3]{ogata1988} and supplement the spatial diagnostics discussed
by Reinhart.

\subsection*{The meaning of location}

Even if the data provided the georeferenced place of residence of each
patient, would that be a suitable proxy for the ``epicentre''?
It may neither be the place where the individual initially became exposed nor
the location receiving the highest triggering rate during the infectious period.
Nevertheless, it is probably the best available proxy.
A more realistic triggering function would obviously
need to employ social contacts rather than spatial displacement.
This is possible in the multivariate models for $\lambda_i(t)$ above but
not for $\lambda(s,t)$, as there is no mapping of locations $s \in X$ to
contact rates. Using a spatio-temporal point process
model for human infections thus entails the assumption that geographic distance
reflects interaction good enough, which is (at least) supported by the
findings of \citet{brockmann.etal2006} and \citet{read.etal2014}.

\subsection*{Underreporting}

Public health surveillance data suffer from considerable
underreporting \citep{gibbons.etal2014}. The consequence is that the
self-exciting model component will be underestimated while the background
process might partially capture cases caused by unobserved sources. This is
similar to the boundary effects discussed in the review.
Indeed, there \emph{is} a background process ``producing new cases
from nowhere'', meaning immigration of infectives from outside the observation
region (e.g., sick tourists or contaminated food), or via antigenic drifts.
To identify such events, stochastic declustering is also of interest in
infectious disease epidemiology, but is less
useful in practice because of the biases from underreporting.

A similar limitation holds for a key epidemiological parameter, the basic
reproduction number $R_0$, estimated as the space-time integral of
the triggering function.
Underreporting and implemented control measures imply
that this estimate is only a lower bound for the \emph{effective}
reproduction number. 
So yes, self-exciting models of infectious disease spread do require careful
interpretation, especially since pathogens in humans are not nearly as well
observable as earthquakes.


\section*{Software}

In synthesizing estimation and inference techniques, the review covers
relevant topics for the analysis of spatio-temporal point patterns from
epidemic phenomena.
I found one crucial aspect to be missing though: software. 
Providing implementations of statistical methods or at least
the code for the specific analysis at hand is essential for scientific
progress today, as it enables others to reproduce the findings and use the
described approaches in their own data-analysis pipelines.

Unsurprisingly, most publically available implementations of self-exciting
point process models are related to the ETAS model.
Several implementations exist for estimating purely temporal versions,
e.g., the Fortran code \code{etas\_solve} by \citet{kasahara.etal2016}, and
the \textsf{R} packages
\pkg{SAPP} \citep{R:SAPP},
\pkg{PtProcess} \citep{harte2010},
and \pkg{bayesianETAS} \citep[see Section~3.5 of the review]{R:bayesianETAS}.
A general-purpose implementation to estimate and simulate purely spatial
cluster process models is provided in the \textsf{R} package
\pkg{spatstat} \citep{baddeley.turner2005}.
The \pkg{ETAS} package \citep{R:ETAS} provides access to
a \textsf{C}/\textsf{C++} port of Zhuang's Fortran routines 
for stochastic declustering in spatio-temporal ETAS models.
There are two sophisticated software packages, which support both
temporal and spatio-temporal ETAS models: \code{SEDA} \citep{lombardi2017}
is a Matlab-based GUI (currently documented to require Mac OS)
for Fortran routines employing
simulated annealing for maximum likelihood estimation, and \pkg{etasFLP}
\citep{adelfio.chiodi2015} is an \textsf{R} package using the
estimation approach described in Section~3.2.2 of the review.


In principle, these ETAS packages could also be used for non-seismological
applications. However, they often do not allow for different parametric
forms of the triggering function, and the modified Omori formula is not
necessarily applicable in other contexts.
For instance, different formulations have been used in crime (Section~4.2)
and epidemic (Section~4.3) forecasting.
At least for epidemiological models, the \textsf{R} package
\pkg{surveillance} \citep{meyer.etal2017} fills the gap. Apart from the
multivariate model of \citet{hoehle2009},
it can also estimate and simulate the spatio-temporal model of
\citet{meyer.etal2012} mentioned in the review.
Various spatial triggering functions are supported,
including Gaussian, power law, student, and (piecewise) constant kernels
(custom forms are possible as well, but will usually be much slower to estimate).
A Newton-type optimizer with analytical derivatives is used to maximize
the log-likelihood. Efforts have been made to avoid vague approximations
of the contained integrals $\int_X f(s-s_i) \,\mathrm{d}s$ over the polygonal
observation region $X$.
Assuming all these integrals to equal 1 is inappropriate for events close to the
boundary and for heavy-tailed kernels in general.
So we compute these integrals, but use an efficient cubature method for
isotropic spatial interaction functions $f$, which only requires one-dimensional
numerical integration (see \citealp[Supplement~B]{meyer.held2014}, and the
\textsf{C} implementation available via the \textsf{R} package \pkg{polyCub}).

\section*{Closing comment}

I hope that Reinhart's review will be 
as infectious as its content and trigger further applications of such
models to epidemic phenomena.
Readily available, well documented, open source software facilitates this
process.

\bibliographystyle{imsart-nameyear}
\bibliography{references}

\end{document}